\documentclass[12pt]{article}
\usepackage{amssymb}
\usepackage{amsmath}
\allowdisplaybreaks
\usepackage{blkarray}
\usepackage{cite}
\usepackage{mathtools}
\usepackage{bbm}

\numberwithin{equation}{section}

\newcommand{\be}{\begin{equation}}
\newcommand{\ee}{\end{equation}}
\newcommand{\non}{\nonumber}

\newcommand{\tr}{\mathop{\rm tr}\nolimits}

\begin{document}

\begin{titlepage}
\strut\hfill UMTG--307
\vspace{.5in}
\begin{center}

\LARGE The $A_m^{(1)}$ Q-system\\ 
\vspace{1in}
\large Rafael I. Nepomechie\footnote{nepomechie@miami.edu}\\
Physics Department, P.O. Box 248046\\
University of Miami, Coral Gables, FL 33124\\[0.8in]
\end{center}

\vspace{.5in}

\begin{abstract}
We propose a Q-system for the $A_m^{(1)}$ quantum integrable spin chain.  We
also find compact determinant expressions for all the Q-functions,
both for the rational and trigonometric cases.  
\end{abstract}

\end{titlepage}

\setcounter{footnote}{0}

\section{Introduction}\label{sec:intro}

Q-systems provide an efficient way of solving Bethe equations
corresponding to quantum integrable models.  Such Q-systems were first
introduced for \emph{rational} Bethe equations (corresponding to
isotropic integrable models) in \cite{Marboe:2016yyn} (see also 
\cite{Kazakov:2007fy}), and have since
been exploited in e.g. \cite{Marboe:2017dmb, Basso:2017khq,
Suzuki:2017ipd, Ryan:2018fyo, Coronado:2018ypq, Jacobsen:2018pjt,
Bajnok:2020xoz}.\footnote{T-systems without spectral parameter are
also called Q-systems \cite{Kuniba:2010ir}, and should not be confused
with the subject of this paper. After this work was completed, we 
became aware that similar relations were investigated in \cite{Mukhin2002SolutionsTT,
Mukhin2006QuasipolynomialsAT, 2013arXiv1303.1578M}; the results of 
\cite{Mukhin2002SolutionsTT} were extended to the trigonometric case 
in \cite{2013JPhCS.411a2020L}.} Generalizations to rank-1
\emph{trigonometric} Bethe equations (corresponding to anisotropic
integrable models) were recently formulated in \cite{Bajnok:2019zub,
Nepomechie:2019gqt}.  One of the principal aims of this work is to
formulate trigonometric Q-systems for higher rank.  We also solve the
rank-$m$ Q-systems (for both the rational and trigonometric cases) in
terms of determinants of a set of $m+1$ functions, which are
generalizations of functions introduced by Pronko and Stroganov to
describe integrable spin chains with $SU(2)$ \cite{Pronko:1998xa} and
$SU(3)$ \cite{Pronko:1999gh} symmetry.  (Throughout this paper, $m=1,
2, \ldots$.)

We first consider the $SU(m+1)$-invariant integrable spin chain in
Section \ref{sec:SU}.  
After briefly reviewing its Bethe ansatz
solution and Q-system \cite{Marboe:2016yyn}, we show that all the
Q-functions can be expressed in terms of compact determinant
expressions of $m+1$ functions $F_{0}, \ldots, F_{m}$
(\ref{Qm})-(\ref{Q0}).  The proof relies on a Pl\"ucker identity
\cite{Krichever:1996qd}.  We then consider the $A_{m}^{(1)}$ integrable spin
chain, which is a q-deformation of the $SU(m+1)$-invariant model, in
Section \ref{sec:Am}.  We propose its Q-system
(\ref{QQAa})-(\ref{QQAb}), and again obtain determinant expressions for all
the Q-functions (\ref{QjA}).  We close with a brief conclusion and a list of some
interesting remaining open problems in Section \ref{sec:conclusion}.

\section{The $SU(m+1)$-invariant spin chain}\label{sec:SU}

We begin by briefly reviewing the closed $SU(m+1)$-invariant integrable quantum spin chain of 
length $N$ with periodic boundary conditions and with ``spins'' in 
the vector ($(m+1)$-dimensional) representation of $SU(m+1)$ at each site, see e.g. \cite{Kulish:1979cr, 
deVega:1989}. The R-matrix (solution of the Yang-Baxter equation)
is given by the $(m+1)^{2} \times (m+1)^{2}$ matrix
\be
\mathbb{R}(u) = u \mathbb{I} + i \mathbb{P} \,,
\ee
where $\mathbb{I}$ is the identity matrix, and $\mathbb{P}$ is the 
permutation matrix, which is given in terms of the elementary 
$(m+1) \times (m+1)$ matrices $e_{a b}$ by
\be
\mathbb{P} =\sum_{a, b = 1}^{m+1} e_{a b} \otimes e_{b a} \,, \qquad 
(e_{a b})_{ij} = \delta_{a,i}\, \delta_{b,j} \,.
\label{perm}
\ee
The transfer matrix $\mathbb{T}(u)$, which is defined by
\be
\mathbb{T}(u) = \tr_{0} \mathbb{R}_{01}(u)\, 
\mathbb{R}_{02}(u)\dots\mathbb{R}_{0N}(u) \,,
\label{transfer}
\ee
satisfies the commutativity property
\be
\left[ \mathbb{T}(u)\,, \mathbb{T}(v) \right] = 0 \,.
\ee
The corresponding spin-chain Hamiltonian is proportional to 
$\frac{d}{du}\left(\log \mathbb{T}(u)\right)\Big\vert_{u=0}$,
up to an additive constant. 

The eigenvalues $T(u)$ of the transfer matrix can be expressed in 
terms of Bethe roots $\{ u_{j,k} \}$ where $k=1, \ldots, M_{j}$
and $j = 1, \ldots, m$, which satisfy the following set of Bethe 
equations
\begin{align}
\left(\frac{u_{1,k}+\frac{i}{2}}{u_{1,k}-\frac{i}{2}}\right)^{N} &=
\prod_{l=1; l \ne k}^{M_{1}} 
\frac{u_{1,k}-u_{1,l}+i}{u_{1,k}-u_{1,l}-i}
\prod_{l=1}^{M_{2}} 
\frac{u_{1,k}-u_{2,l}-\frac{i}{2}}{u_{1,k}-u_{2,l}+\frac{i}{2}}
\,, \quad k = 1, \ldots, M_{1}\,, \label{BESU1}\\
1 &=
\prod_{l=1; l \ne k}^{M_{j}} 
\frac{u_{j,k}-u_{j,l}+i}{u_{j,k}-u_{j,l}-i}
\prod_{l=1}^{M_{j+1}} 
\frac{u_{j,k}-u_{j+1,l}-\frac{i}{2}}{u_{j,k}-u_{j+1,l}+\frac{i}{2}}
\prod_{l=1}^{M_{j-1}}\frac{u_{j,k}-u_{j-1,l}-\frac{i}{2}}
{u_{j,k}-u_{j-1,l}+\frac{i}{2}}\,, \non \\
&\qquad\qquad k = 1, \ldots, M_{j}\,,
\qquad j = 2, \ldots, m-1\,, \label{BESUj} \\
1 &=
\prod_{l=1; l \ne k}^{M_{m}} 
\frac{u_{m,k}-u_{m,l}+i}{u_{m,k}-u_{m,l}-i}
\prod_{l=1}^{M_{m-1}}\frac{u_{m,k}-u_{m-1,l}-\frac{i}{2}}
{u_{m,k}-u_{m-1,l}+\frac{i}{2}}\,, \quad k = 1, \ldots, M_{m} 
\,. \label{BESUm}
\end{align}
We define polynomials $Q_{1}(u), \ldots, Q_{m}(u)$ by
\be
Q_{j}(u) = \prod_{k=1}^{M_{j}}(u-u_{j,k}) \,, \qquad j = 1, \ldots, m 
\,, \label{Qfuncs}
\ee
so that their zeros are given by corresponding Bethe roots.

\subsection{The $SU(m+1)$ Q-system}\label{sec:QSU}

The QQ-relations for the $SU(m+1)$ Q-system are given by \cite{Marboe:2016yyn}
\begin{align}
Q_{j,n}(u)\, Q_{j+1,n-1}(u) & \propto Q_{j+1,n}^{+}(u)\, Q_{j,n-1}^{-}(u) - 
Q_{j+1,n}^{-}(u)\, Q_{j,n-1}^{+}(u) \,, \non \\
& \qquad\qquad j=0, 1, \ldots, m\,, \qquad n=1, 2, \ldots \,,  \label{QQSUa}
\end{align}
where $f^{\pm}(u)=f(u\pm\frac{i}{2})$, and 
\begin{align}
Q_{0,0}(u) &= u^{N} \,, \non\\
Q_{j,0}(u) &= Q_{j}(u) \,, \quad j = 1, \ldots, m\,, \non\\
Q_{m+1,0}(u) &= 1 \,, \label{QQSUb}
\end{align}
where $Q_{j}(u)$ are defined in (\ref{Qfuncs}). The 
nontrivial Q-functions are in fact defined on a Young diagram with 
$N$ boxes,
on whose boundary the Q-functions (including $Q_{m+1,0}$) are set to 1 \cite{Marboe:2016yyn}.

Let us verify that this Q-system indeed leads to the Bethe 
equations. Setting $n=1$ in (\ref{QQSUa}) and using 
(\ref{QQSUb}), we obtain
\begin{align}
Q_{j,1}(u)\, Q_{j+1}(u) & \propto Q_{j+1,1}^{+}(u)\, Q_{j}^{-}(u) - 
Q_{j+1,1}^{-}(u)\, Q_{j}^{+}(u) \,, \quad j=1, \ldots, m-1  \,.
\label{QQBEa}
\end{align}
Shifting $j \mapsto j-1$ in (\ref{QQBEa}) and then setting $u=u_{j,k}$,
we obtain
\be
0 = Q_{j,1}^{+}(u_{j,k})\, Q_{j-1}^{-}(u_{j,k}) - 
Q_{j,1}^{-}(u_{j,k})\, Q_{j-1}^{+}(u_{j,k}) \,, \label{QQBEb}
\ee
since $Q_{j}(u_{j,k})=0$.
Shifting $u \mapsto u \pm \frac{i}{2}$ in (\ref{QQBEa}) and then 
setting $u=u_{j,k}$, we obtain the pair of equations
\begin{align}
Q_{j,1}^{+}(u_{j,k})\, Q_{j+1}^{+}(u_{j,k}) &\propto -Q_{j+1,1}(u_{j,k})\, 
Q_{j}^{++}(u_{j,k}) \,, \non \\
Q_{j,1}^{-}(u_{j,k})\, Q_{j+1}^{-}(u_{j,k}) &\propto Q_{j+1,1}(u_{j,k})\, 
Q_{j}^{--}(u_{j,k}) \,. \label{QQBEc}
\end{align}
Using (\ref{QQBEc}) to eliminate $Q_{j,1}^{\pm}(u_{j,k})$ in 
(\ref{QQBEb}), we arrive at the relations
\be
\frac{Q_{j}^{++}(u_{j,k})}{Q_{j+1}^{+}(u_{j,k})}Q_{j-1}^{-}(u_{j,k}) 
= - 
\frac{Q_{j}^{--}(u_{j,k})}{Q_{j+1}^{-}(u_{j,k})}Q_{j-1}^{+}(u_{j,k})\,, 
\quad j = 2, \ldots, m-1 \,, \label{QQBEd}
\ee
which is equivalent to the Bethe equations (\ref{BESUj}). 

To get the first Bethe equation (\ref{BESU1}), we start
from (\ref{QQSUa}) with $n=1$ and $j=0$
\be
Q_{0,1}(u)\, Q_{1}(u)  \propto Q_{1,1}^{+}(u)\, Q_{0,0}^{-}(u) - 
Q_{1,1}^{-}(u)\, Q_{0,0}^{+}(u) \,.
\ee
Evaluating this relation at $u=u_{1,k}$ gives
\be
0 = Q_{1,1}^{+}(u_{1,k})\, Q_{0,0}^{-}(u_{1,k}) - 
Q_{1,1}^{-}(u_{1,k})\, Q_{0,0}^{+}(u_{1,k}) \,. \label{QQBEe}
\ee
Using (\ref{QQBEc}) with $j=1$ to eliminate $Q_{1,1}^{\pm}(u_{1,k})$ 
in (\ref{QQBEe}),
we arrive at (\ref{QQBEd}) with $j=1$, except with $Q_{0}$ replaced 
by $Q_{0,0}$, which is equivalent to the Bethe equation (\ref{BESU1}).

To get the final Bethe equation (\ref{BESUm}), we start
from (\ref{QQSUa}) with $n=1$ and $j=m$ \footnote{Eq. 
(\ref{QQBEf}) holds if $M_{m}>1$; however, $Q_{m,1}=1$ if 
$M_{m}=1$, which is consistent with (\ref{Qm}).}
\be
Q_{m,1}(u)  \propto Q_{m+1,1}^{+}(u)\, Q_{m}^{-}(u) - 
Q_{m+1,1}^{-}(u)\, Q_{m}^{+}(u) \,. \label{QQBEf}
\ee
Shifting $u \mapsto u \pm \frac{i}{2}$ in (\ref{QQBEf}) and then 
setting $u=u_{m,k}$, we obtain (\ref{QQBEc}) with $j=m$, except with
$Q_{m+1}$ replaced by 1. Using these relations to eliminate 
$Q_{m,1}^{\pm}(u_{m,k})$ in (\ref{QQBEb}) with $j=m$, we arrive at 
(\ref{QQBEd}) with $j=m$, except with $Q_{m+1}$ replaced 
by 1, which is indeed equivalent to the Bethe equation (\ref{BESUm}).

\subsection{Determinant representation for all the Q-functions}\label{sec:detSU}

We now show that the Q-system (\ref{QQSUa})-(\ref{QQSUb}) can be solved in terms of 
a set of $m+1$ functions $F_{0}(u), \ldots, F_{m}(u)$, 
whose interpretation will be discussed later.
Explicitly, all the Q-functions can be expressed in terms of 
determinants as follows \footnote{Similar formulas were found for 
bosonic spin chains in \cite{Mukhin2002SolutionsTT}, and for 
supersymmetric spin chains in \cite{Tsuboi:2009ud, Tsuboi:2011iz}.
}
\begin{align}
	Q_{m,n} &= F_{0}^{(n)} \,,  \label{Qm} \\
	Q_{m-1,n} &= \begin{vmatrix}
	                F_{0}^{(n)+} & F_{0}^{(n)-}\\
					F_{1}^{(n)+} & F_{1}^{(n)-}
					\end{vmatrix}_{2 \times 2}\,, \label{Qm1} \\
		     & \vdots \non \\
	Q_{j,n} &=	
	\begin{vmatrix}
	      F_{0}^{(n)[m-j]} & F_{0}^{(n)[m-j-2]} &\cdots & F_{0}^{(n)[j-m]}\\
		  F_{1}^{(n)[m-j]} & F_{1}^{(n)[m-j-2]} &\cdots & F_{1}^{(n)[j-m]}\\
		  \vdots           &  \vdots            &       &  \vdots  \\
		  F_{m-j}^{(n)[m-j]} & F_{m-j}^{(n)[m-j-2]} &\cdots & F_{m-j}^{(n)[j-m]}
		  \end{vmatrix}_{(m+1-j) \times (m+1-j)} \,, \label{Qj} \\
		  & \vdots \non \\
	Q_{0,n} &=	
	\begin{vmatrix}
	      F_{0}^{(n)[m]} & F_{0}^{(n)[m-2]} &\cdots & F_{0}^{(n)[-m]}\\	  
		  F_{1}^{(n)[m]} & F_{1}^{(n)[m-2]} &\cdots & F_{1}^{(n)[-m]}\\
		  \vdots           &  \vdots            &       &  \vdots  \\
		  F_{m}^{(n)[m]} & F_{m}^{(n)[m-2]} &\cdots & F_{m}^{(n)[-m]}
		  \end{vmatrix}_{(m+1) \times (m+1)} \,,	\label{Q0}		
\end{align}
where $n=0, 1, \ldots$.
Throughout Section \ref{sec:SU}, we use the notation $f^{(n)}(u)$ to 
denote the $n^{th}$ discrete derivative of any function $f(u)$, i.e.
\begin{align}
f^{(n)}(u) &=f^{(n-1)+}(u) - f^{(n-1)-}(u) \non \\
           &= f^{(n-1)}(u+\tfrac{i}{2}) - 
f^{(n-1)}(u-\tfrac{i}{2})\,, \qquad n = 1, 2, \ldots , \label{discderiv}
\end{align}
with $f^{(0)}(u)=f(u)$. Moreover, we use the notation $f^{[k]}(u)$ 
to denote a $k$-fold shift by $\tfrac{i}{2}$, i.e.
\be
f^{[k]}(u) = f(u + k \tfrac{i}{2}) \,.
\label{kshift}
\ee
Hence, $f^{[1]}=f^{+}$ and $f^{[-1]}=f^{-}$, etc. Note that 
(\ref{QQSUb}) and (\ref{Qm}) imply that 
\be
Q_{m} = Q_{m,0} =  F_{0} \,. \label{QF}
\ee

An important consequence of the result (\ref{Qm})-(\ref{Q0}) is that
the functions $F_{0}(u), \ldots, F_{m}(u)$ are polynomials in $u$ if 
and only if all the Q-functions are polynomials in $u$.

In order to show that the expressions (\ref{Qm})-(\ref{Q0}) indeed satisfy 
the QQ-relations (\ref{QQSUa}), we make use of Pl\"ucker identities
\cite{Krichever:1996qd}, which we first briefly review. Let $X$ 
denote a {\em rectangular} matrix with $r$ rows and $c$ columns, with 
$c>r$,
\be
X = \begin{pmatrix}
X_{1,1} & X_{1,2} & \cdots & X_{1,c} \\
X_{2,1} & X_{2,2} & \cdots & X_{2,c} \\
\vdots  & \vdots  & \cdots & \vdots  \\
X_{r,1} & X_{r,2} & \cdots & X_{r,c} 
\end{pmatrix} \,.
\ee
Furthermore, let the symbol $(i_{1}, i_{2}, \ldots, i_{r})$ denote the 
determinant of the square matrix formed by the $r$ columns $i_{1}, 
i_{2}, \ldots, i_{r}$ of $X$
\be
(i_{1}, i_{2}, \ldots, i_{r}) =
\begin{vmatrix}
X_{1,i_1} & X_{1,i_2} & \cdots & X_{1,i_{r}} \\
X_{2,i_1} & X_{2,i_2} & \cdots & X_{2,i_{r}} \\
\vdots  & \vdots  & \cdots & \vdots  \\
X_{r,i_1} & X_{r,i_2} & \cdots & X_{r,i_{r}} 
\end{vmatrix} \,,
\ee
which is antisymmetric in all indices. The particular Pl\"ucker 
identity that we need is \cite{Krichever:1996qd}
\begin{align}
(i_{1}, i_{2}, k_{3}, \ldots, k_{r})\, (j_{1}, j_{2}, k_{3}, \ldots, 
k_{r}) &= 
(j_{1}, i_{2}, k_{3}, \ldots, k_{r})\, (i_{1}, j_{2}, k_{3}, \ldots, 
k_{r}) \non \\
&+
(j_{2}, i_{2}, k_{3}, \ldots, k_{r})\, (j_{1}, i_{1}, k_{3}, \ldots, 
k_{r}) \,,
\label{Plucker}
\end{align}
where all indices take values in $\{1, 2, \ldots, c\}$.

For the problem at hand, we choose $X$ to be a rectangular matrix
with $r=m+1-j$ and $c=m+3-j$ (where $j=0, 1, \ldots, m$) given by
\be
X = 
\begin{blockarray}{cccccc}
1 & 2 & \cdots & m+1-j & m+2-j & m+3-j \\[0.3cm]
\begin{block}{(cccccc)}
F_{0}^{(n)[m-j]} & F_{0}^{(n)[m-j-2]} &\cdots & F_{0}^{(n)[j-m]} & 
F_{0}^{(n-1)[m-j-1]} & 0\\
F_{1}^{(n)[m-j]} & F_{1}^{(n)[m-j-2]} &\cdots & F_{1}^{(n)[j-m]} &
F_{1}^{(n-1)[m-j-1]} & 0\\
\vdots           &  \vdots            &       &  \vdots  & \vdots & 
\vdots \\
F_{m-j-1}^{(n)[m-j]} & F_{m-j-1}^{(n)[m-j-2]} &\cdots & F_{m-j-1}^{(n)[j-m]} & 
F_{m-j-1}^{(n-1)[m-j-1]} & 0 \\[0.1cm]
F_{m-j}^{(n)[m-j]} & F_{m-j}^{(n)[m-j-2]} &\cdots & F_{m-j}^{(n)[j-m]} &
F_{m-j}^{(n-1)[m-j-1]} & 1 \\
\end{block} 
\end{blockarray} 
\label{Xj} \,.
\ee 
Moreover, we choose the indices in (\ref{Plucker}) as follows
\be
i_{1}=1\,, \quad i_{2}=m+1-j\,, \quad j_{1} = m+2-j\,, \quad j_{2} = m+3-j\,, 
\label{choice1}
\ee 
and 
\be
k_{l} = l-1 \quad \mbox{ for } \quad  l = 3, 4, \ldots, m+1-j \,.
\label{choice2}
\ee
With the choices (\ref{choice1})-(\ref{choice2}), the Pl\"ucker
identity (\ref{Plucker}) reads
\begin{align}
& (1, 2, \ldots, m-j, m+1-j)\, (2, \ldots, m-j, m+2-j, m+3-j)  \non\\
& \propto (2, \ldots, m-j, m+1-j, m+2-j)\, (1, 2, \ldots, m-j, m+3-j) \non\\
&\quad -(2, \ldots, m-j, m+1-j, m+3-j)\, (1, 2, \ldots, m-j, m+2-j)\,,
\label{Plucker2}
\end{align}
where we have made use of the antisymmetry properties of the symbols. 
Using the following identifications
\begin{align}
	(1, 2, \ldots, m-j, m+1-j) & = Q_{j,n} \,, \label{id1} \\
	(2, \ldots, m-j, m+2-j, m+3-j) & = Q_{j+1,n-1} \,, \label{id2}\\
	(2, \ldots, m-j, m+1-j, m+2-j) & = Q_{j,n-1}^{-} \,, \label{id3} \\
	(1, 2, \ldots, m-j, m+3-j) & = Q_{j+1,n}^{+} \,, \label{id4} \\
	(2, \ldots, m-j, m+1-j, m+3-j) & = Q_{j+1,n}^{-} \,, \label{id5} \\
	(1, 2, \ldots, m-j, m+2-j)  & = Q_{j,n-1}^{+} \,, \label{id6}
\end{align}	
we immediately obtain from the identity (\ref{Plucker2}) the
QQ-relations (\ref{QQSUa}).  The first (\ref{id1}), fourth (\ref{id4})
and fifth (\ref{id5}) relations follow directly from (\ref{Qj}) and
the expression (\ref{Xj}) for $X$; while the second (\ref{id2}), third
(\ref{id3}) and sixth (\ref{id6}) relations, which involve Q-functions
with $n-1$ instead of $n$, require also (\ref{discderiv}).

The determinant expressions (\ref{Qm})-(\ref{Q0}) for the 
Q-functions and their proof constitute some of the main results of this 
paper. We remark that the result (\ref{Qj}) with $n=0$ is similar to Eq. (9.21)
in \cite{Kuniba:2010ir}.

In order to understand how to interpret the functions $F_{0}(u),
\ldots, F_{m}(u)$, it is helpful to begin by analyzing the simplest
cases $m= 1, 2$.

\paragraph{$m=1$}

For the $SU(2)$ case ($m=1$), the results (\ref{Qm})-(\ref{Q0}) 
reduce to
\be
	Q_{1,n} = F_{0}^{(n)} \,,  \qquad 
	Q_{0,n} = \begin{vmatrix}
	                F_{0}^{(n)+} & F_{0}^{(n)-}\\
					F_{1}^{(n)+} & F_{1}^{(n)-}
			  \end{vmatrix} = F_{0}^{(n)+}\, F_{1}^{(n)-} - 
			  F_{0}^{(n)-}\, F_{1}^{(n)+} \,. \label{exSU2}
\ee
In view of (\ref{QF}), we can identify $F_{0}(u)$ as the 
``fundamental'' 
Q-function $F_{0}(u) = Q_{1}(u) \equiv Q(u)$. Moreover, 
we can identify $F_{1}(u)$ as the ``dual'' Q-function, which is 
denoted by $P(u)$ in \cite{Pronko:1998xa} and \cite{Bajnok:2019zub}, 
see also \cite{Nepomechie:2019gqt}.
Indeed, with these identifications, (\ref{exSU2}) coincides with Eq. 
(2.23) in \cite{Bajnok:2019zub}. Note that (\ref{exSU2}) with $n=0$ 
implies that
\be
Q_{0,0} = F_{0}^{+}\, F_{1}^{-} - F_{0}^{-}\, F_{1}^{+} \,,
\ee
which we recognize as the important discrete Wronskian relation in \cite{Pronko:1998xa}.

It was proved in \cite{Mukhin:2009, Tarasov:2018} that polynomiality
of $Q$ and $P$ (i.e., $F_{0}$ and $F_{1}$) is equivalent to the
admissibility of the Bethe roots.  It follows from (\ref{exSU2}) that
polynomiality of the Q-system is equivalent to the admissibility of
the Bethe roots, as already noted in \cite{Marboe:2016yyn}.\footnote{We remind the reader that some
solutions of the Bethe equations, most notably the so-called
unphysical singular solutions, do not lead to eigenvalues and
eigenvectors of the transfer matrix, see e.g. \cite{Bajnok:2019zub}
and references therein.  Here we call \emph{admissible} those
solutions of the Bethe equations that \emph{do} give rise to genuine
eigenvalues and eigenvectors of the transfer matrix.}

\paragraph{$m=2$}

For the $SU(3)$ case ($m=2$), the results (\ref{Qm})-(\ref{Q0}) 
reduce to
\begin{align}
	Q_{2,n} &= F_{0}^{(n)} \,,   \\
	Q_{1,n} &= \begin{vmatrix}
	                F_{0}^{(n)+} & F_{0}^{(n)-}\\
					F_{1}^{(n)+} & F_{1}^{(n)-}
					\end{vmatrix}\,,  \\
	Q_{0,n} &=	
	\begin{vmatrix}
	      F_{0}^{(n)++} & F_{0}^{(n)} & F_{0}^{(n)--}\\	  
		  F_{1}^{(n)++} & F_{1}^{(n)} & F_{1}^{(n)--}\\
		  F_{2}^{(n)++} & F_{2}^{(n)} & F_{2}^{(n)--}
		  \end{vmatrix} \,.		\label{exSU3}
\end{align}
The functions $F_{0}, F_{1}, F_{2}$ correspond to the functions $P, 
Q, R$ in \cite{Pronko:1999gh}. In particular, (\ref{exSU3}) with $n=0$ 
can be recognized as Eq. (4) in \cite{Pronko:1999gh}.

\paragraph{General $m$}

We now see that the functions $F_{0}(u), \ldots, F_{m}(u)$ are
generalizations of the functions introduced by Pronko and Stroganov to
describe integrable spin chains with $SU(2)$ \cite{Pronko:1998xa} and
$SU(3)$ \cite{Pronko:1999gh} symmetry.  These functions satisfy a
generalized discrete Wronskian relation given by (\ref{Q0}) with
$n=0$. We conjecture, generalizing the $m=1$ result of
\cite{Mukhin:2009, Tarasov:2018}, that polynomiality of $F_{0}(u), \ldots, F_{m}(u)$
(and hence, by (\ref{Qm})-(\ref{Q0}), polynomiality of all the 
Q-functions) is equivalent to the admissibility of the Bethe roots $\{ u_{j,k} \}$.

\section{The $A_{m}^{(1)}$ spin chain}\label{sec:Am}

We turn now to the closed $A_{m}^{(1)}$ spin chain with periodic boundary conditions, 
which is a q-deformation of the $SU(m+1)$-invariant model 
considered in Section \ref{sec:SU}. The R-matrix is now given by (see 
e.g. \cite{Jimbo:1985ua}, with $x=e^{2u}$ and $k=e^{-\eta}$)
\begin{align}
\mathbb{R}(u) &= \frac{1}{2}e^{\eta-u}\Bigg\{ 
(e^{2u}-e^{-2\eta})\sum_{a=1}^{m+1} e_{a a} \otimes e_{a a} + 
e^{-\eta}(e^{2u}-1)\sum_{a\ne b} e_{a a} \otimes e_{b b} \non\\
& - (e^{-2\eta}-1)\left(\sum_{a<b} + e^{2u} \sum_{a>b}\right) e_{a b} 
\otimes e_{b a} \Bigg\} \,,
\label{RAm}
\end{align}
where $\eta$ is the anisotropy parameter.
The transfer matrix $\mathbb{T}(u)$ is again given by (\ref{transfer})
\be
\mathbb{T}(u) = \tr_{0} \mathbb{R}_{01}(u)\, 
\mathbb{R}_{02}(u)\dots\mathbb{R}_{0N}(u) \,.
\label{transferAm}
\ee
The Bethe equations are now given by
\begin{align}
\left(\frac{\sinh(u_{1,k}+\frac{\eta}{2})}{\sinh(u_{1,k}-\frac{\eta}{2})}\right)^{N} &=
\prod_{l=1; l \ne k}^{M_{1}} 
\frac{\sinh(u_{1,k}-u_{1,l}+\eta)}{\sinh(u_{1,k}-u_{1,l}-\eta)}
\prod_{l=1}^{M_{2}} 
\frac{\sinh(u_{1,k}-u_{2,l}-\frac{\eta}{2})}{\sinh(u_{1,k}-u_{2,l}+\frac{\eta}{2})} 
\,, \quad k = 1, \ldots, M_{1}\,, \label{BEA1}\\
1 &=
\prod_{l=1; l \ne k}^{M_{j}} 
\frac{\sinh(u_{j,k}-u_{j,l}+\eta)}{\sinh(u_{j,k}-u_{j,l}-\eta)}
\prod_{l=1}^{M_{j+1}} 
\frac{\sinh(u_{j,k}-u_{j+1,l}-\frac{\eta}{2})}{\sinh(u_{j,k}-u_{j+1,l}+\frac{\eta}{2})}\non \\
& \qquad \times
\prod_{l=1}^{M_{j-1}}\frac{\sinh(u_{j,k}-u_{j-1,l}-\frac{\eta}{2})}
{\sinh(u_{j,k}-u_{j-1,l}+\frac{\eta}{2})}\,, \quad k = 1, \ldots, M_{j}\,,
\quad j = 2, \ldots, m-1\,, \label{BEAj} \\
1 &=
\prod_{l=1; l \ne k}^{M_{m}} 
\frac{\sinh(u_{m,k}-u_{m,l}+\eta)}{\sinh(u_{m,k}-u_{m,l}-\eta)}
\prod_{l=1}^{M_{m-1}}\frac{\sinh(u_{m,k}-u_{m-1,l}-\frac{\eta}{2})}
{\sinh(u_{m,k}-u_{m-1,l}+\frac{\eta}{2})}\,, \quad k = 1, \ldots, M_{m} 
\,, \label{BEAm}
\end{align}
and we now define functions $Q_{1}(u), \ldots, Q_{m}(u)$  by
\be
Q_{j}(u) = \prod_{k=1}^{M_{j}}\sinh(u-u_{j,k}) \,, \qquad j = 1, \ldots, m 
\,, \label{QfuncsA}
\ee
which are polynomials in $t\equiv e^{u}$ and $t^{-1}$.

\subsection{The $A_{m}^{(1)}$ Q-system}\label{sec:QA}

We propose that the $A_{m}^{(1)}$ spin chain has the same 
QQ-relations as the isotropic case (\ref{QQSUa}), namely
\begin{align}
Q_{j,n}(u)\, Q_{j+1,n-1}(u) & \propto Q_{j+1,n}^{+}(u)\, Q_{j,n-1}^{-}(u) - 
Q_{j+1,n}^{-}(u)\, Q_{j,n-1}^{+}(u) \,, \non \\
& \qquad\qquad j=0, 1, \ldots, m\,, \qquad n=1, 2, \ldots \,,  
\label{QQAa}
\end{align}
but where now $f^{\pm}(u)=f(u\pm\frac{\eta}{2})$. Moreover, 
\begin{align}
Q_{0,0}(u) &= \sinh^{N}(u) \,, \non\\
Q_{j,0}(u) &= Q_{j}(u) \,, \quad j = 1, \ldots, m\,, \non\\
Q_{m+1,0}(u) &= 1 \,, \label{QQAb}
\end{align}
where the functions $Q_{j}(u)$ are defined in (\ref{QfuncsA}).

We can easily verify that this Q-system indeed leads to the
$A_{m}^{(1)}$ Bethe equations.  Indeed, starting from
(\ref{QQAa})-(\ref{QQAb}) and repeating the steps in Section
\ref{sec:QSU}, we arrive at the Bethe equations
(\ref{BEA1})-(\ref{BEAm}).

As in the isotropic case, the $A_{m}^{(1)}$ Q-system can be solved in terms
of functions $F_{0}(u), \ldots, F_{m}(u)$ by 
\be
Q_{j,n} =	\begin{vmatrix}
	      F_{0}^{(n)[m-j]} & F_{0}^{(n)[m-j-2]} &\cdots & F_{0}^{(n)[j-m]}\\
		  F_{1}^{(n)[m-j]} & F_{1}^{(n)[m-j-2]} &\cdots & F_{1}^{(n)[j-m]}\\
		  \vdots           &  \vdots            &       &  \vdots  \\
		  F_{m-j}^{(n)[m-j]} & F_{m-j}^{(n)[m-j-2]} &\cdots & F_{m-j}^{(n)[j-m]}
		  \end{vmatrix}_{(m+1-j) \times (m+1-j)} \,, \quad j = 0, 1, 
		  \ldots, m\,, \label{QjA} 
\ee
and $n= 0, 1, \ldots$, except now
\begin{align}
f^{(n)}(u) &=f^{(n-1)+}(u) - f^{(n-1)-}(u) \non \\
           &= f^{(n-1)}(u+\tfrac{\eta}{2}) - 
f^{(n-1)}(u-\tfrac{\eta}{2})\,, \qquad n = 1, 2, \ldots, 
\end{align}
and
\be
f^{[k]}(u) = f(u + k \tfrac{\eta}{2}) \,.
\ee
Indeed, the same proof from Section \ref{sec:detSU} carries over to the anisotropic case.

The $A_{m}^{(1)}$ Q-system (\ref{QQAa})-(\ref{QQAb}) and its determinant
representation (\ref{QjA}) constitute the other main results of this
paper. 

We conjecture that polynomiality of $F_{0}$ in 
$t\equiv e^{u}$ and $t^{-1}$, together with
quasi-polynomiality (polynomial plus $\log t$ times a polynomial) of 
$F_{1}, \ldots, F_{m}$, is equivalent to the admissibility of the 
Bethe roots; and is also equivalent to polynomiality of all the 
Q-functions. Evidence supporting this conjecture for the simplest cases $m= 1, 2$ 
is provided below.

\paragraph{$m=1$}

The $A_{1}^{(1)}$ case ($m=1$) corresponds to the spin-1/2 XXZ spin chain,
whose Q-system was recently formulated in \cite{Bajnok:2019zub}.
As in the isotropic case, we identify $F_{0}(u)$ as the
``fundamental'' Q-function $Q(u)$, and $F_{1}(u)$ as the ``dual''
Q-function $P(u)$. The relations (\ref{QjA}) with $m=1$ coincide with 
Eq. (3.13) in \cite{Bajnok:2019zub}.

It was argued in \cite{Bajnok:2019zub} that polynomiality of $Q$,
together with quasi-polynomiality of $P$, is equivalent to the
admissibility of the Bethe roots.  It follows from (\ref{QjA}) with
$m=1$ that polynomiality of the Q-system is equivalent to the
admissibility of the Bethe roots.

\paragraph{$m=2$}

For the $A_{2}^{(1)}$ case ($m=2$), we have verified numerically for small 
values of $N$ that the 
polynomial (in $t\equiv e^{u}$ and $t^{-1}$) solutions of the 
Q-system (\ref{QQAa})-(\ref{QQAb}) give the complete spectrum of the 
transfer matrix $\mathbb{T}(u)$ (\ref{transferAm}). Indeed, we have used this Q-system 
to numerically obtain the admissible Bethe roots for some generic 
value of $\eta$, 
as in the $A_{1}^{(1)}$ case \cite{Bajnok:2019zub}; and we have verified 
that the corresponding eigenvalues $T(u)$ of the transfer matrix computed using
\be
	T(u)=Q_{0,0}^{++}(u)\frac{Q_{1}^{-}(u)}{Q_{1}^{+}(u)}
	+Q_{0,0}(u)\frac{Q_{1}^{[3]}(u)}{Q_{1}^{+}(u)}\frac{Q_{2}(u)}{Q_{2}^{++}(u)}
	+Q_{0,0}(u)\frac{Q_{2}^{[4]}(u)}{Q_{2}^{++}(u)} \label{Teig}
\ee
match with the results obtained by direct diagonalization of $\mathbb{T}(u)$. 

We report in Table \ref{table:A2}, for given values 
of chain length $N$, the numbers of Bethe roots of each type ($M_{1}, 
M_{2}$), the corresponding number of admissible solutions ($n_{M_{1}, M_{2}}$) 
of the Bethe equations (\ref{BEA1})-(\ref{BEAm}) obtained by solving the 
Q-system (\ref{QQAa})-(\ref{QQAb}), and the degeneracies ($d_{M_{1}, M_{2}}$) of the 
corresponding transfer-matrix eigenvalues $T(u)$ (\ref{Teig}) obtained by direct 
diagonalization of $\mathbb{T}(u)$. (We refrain from displaying the 
Bethe roots themselves, which would require much bigger tables.) 
The completeness is demonstrated by the fact (easily confirmed from the 
data in Table \ref{table:A2}) that all $3^{N}$ transfer-matrix eigenvalues are 
accounted for, i.e.
\be
\sum_{M_{1},M_{2}} n_{M_{1},M_{2}}\, d_{M_{1},M_{2}} = 3^{N} \,.
\ee 

\begin{table}[htb]
\small 
\centering
\begin{tabular}{|c|c|c|c|c|}
\hline
$N$ & $M_{1}$ & $M_{2}$ & $n_{M_{1}, M_{2}}$ & $d_{M_{1}, M_{2}}$ \\
\hline
2 & 0 & 0 & 1 & 3 \\
  & 1 & 0 & 2 & 3 \\
\hline
3 & 0 & 0 & 1 & 3 \\
  & 1 & 0 & 3 & 6 \\
  & 2 & 1 & 6 & 1 \\
\hline
4 & 0 & 0 & 1 & 3 \\
  & 1 & 0 & 4 & 6 \\
  & 2 & 0 & 6 & 3 \\
  & 2 & 1 & 12 & 3 \\
\hline
5 & 0 & 0 & 1 & 3 \\
  & 1 & 0 & 5 & 6 \\
  & 2 & 0 & 10 & 6 \\
  & 2 & 1 & 20 & 3 \\
  & 3 & 1 & 30 & 3 \\
\hline
\end{tabular}
\caption{\small Chain length ($N$), numbers of Bethe roots ($M_{1}, 
M_{2}$), number of admissible solutions of the Bethe equations ($n_{M_{1}, M_{2}}$) 
and degeneracies ($d_{M_{1}, M_{2}}$) for the $A_{2}^{(1)}$ spin chain}
\label{table:A2}
\end{table}

\section{Conclusions}\label{sec:conclusion}

We have proposed a Q-system for the $A_{m}^{(1)}$ spin chain
(\ref{QQAa})-(\ref{QQAb}), which provides an efficient way of
obtaining solutions of the trigonometric Bethe equations
(\ref{BEA1})-(\ref{BEAm}).  We have also found compact determinant
expressions for all the Q-functions, both for the rational
(\ref{Qm})-(\ref{Q0}) and trigonometric (\ref{QjA}) cases. In so 
doing, we have established links between Q-systems and the works by 
Kuniba et al. \cite{Kuniba:2010ir} and by Pronko and Stroganov 
\cite{Pronko:1998xa, Pronko:1999gh}.

Several interesting related problems remain to be addressed. The 
fact that all the Q-functions can be expressed in terms of the $m+1$
F-functions $F_{0}, \ldots, F_{m}$ 
suggests that the F-functions are of fundamental importance for the 
$SU(m+1)$ and $A_{m}^{(1)}$ models, and merit further investigation. In particular, 
it would be desirable to have proofs that polynomiality in $u$ (or 
quasi-polynomiality in $t$ and $t^{-1}$ for the trigonometric case) 
of the F-functions is equivalent to the admissibility of the Bethe roots.
For $SU(m|n)$ graded (supersymmetric) spin chains \cite{Kulish:1985bj}, 
Q-systems were also formulated in \cite{Marboe:2016yyn, Kazakov:2007fy}; 
it should be possible to formulate similar determinant
expressions for these Q-systems and for the corresponding q-deformed 
models, and to relate them to results of Tsuboi \cite{Tsuboi:2009ud, Tsuboi:2011iz}.
We restricted here to periodic boundary conditions; it should
be possible to generalize the $A_{m}^{(1)}$ Q-system (and the corresponding
determinant representation) to open diagonal boundary conditions
\cite{Sklyanin:1988yz, deVega:1992zd, deVega:1994sb,Doikou:1998ek}, thereby
generalizing the corresponding rank-1 results
\cite{Bajnok:2019zub,Nepomechie:2019gqt} to higher rank.  Only A-type
Q-systems are so far known; it would be very interesting to construct such
Q-systems for other algebras.

\section*{Acknowledgments}
I am grateful to Z. Bajnok, E. Granet, J. Jacobsen, Y. Jiang and Y. Zhang for 
their collaboration on related earlier projects \cite{Bajnok:2019zub, 
Bajnok:2020xoz}, from which this work arose. I also thank A. Gainutdinov 
for valuable discussions.

% \bibliographystyle{utphys}
% \bibliography{refs}

\begin{thebibliography}{10}

\bibitem{Marboe:2016yyn}
C.~Marboe and D.~Volin, ``{Fast analytic solver of rational Bethe equations},''
  \href{http://dx.doi.org/10.1088/1751-8121/aa6b88}{{\em J. Phys.} {\bfseries
  A50} no.~20, (2017) 204002},
\href{http://arxiv.org/abs/1608.06504}{{\ttfamily arXiv:1608.06504 [math-ph]}}.
%%CITATION = ARXIV:1608.06504;%%.

\bibitem{Kazakov:2007fy}
V.~Kazakov, A.~S. Sorin, and A.~Zabrodin, ``{Supersymmetric Bethe ansatz and
  Baxter equations from discrete Hirota dynamics},''
  \href{http://dx.doi.org/10.1016/j.nuclphysb.2007.06.025}{{\em Nucl. Phys.}
  {\bfseries B790} (2008) 345--413},
\href{http://arxiv.org/abs/hep-th/0703147}{{\ttfamily arXiv:hep-th/0703147
  [HEP-TH]}}.
%%CITATION = HEP-TH/0703147;%%.

\bibitem{Marboe:2017dmb}
C.~Marboe and D.~Volin, ``{The full spectrum of AdS5/CFT4 I: Representation
  theory and one-loop Q-system},''
  \href{http://dx.doi.org/10.1088/1751-8121/aab34a}{{\em J. Phys.} {\bfseries
  A51} no.~16, (2018) 165401},
\href{http://arxiv.org/abs/1701.03704}{{\ttfamily arXiv:1701.03704 [hep-th]}}.
%%CITATION = ARXIV:1701.03704;%%.

\bibitem{Basso:2017khq}
B.~Basso, F.~Coronado, S.~Komatsu, H.~T. Lam, P.~Vieira, and D.-l. Zhong,
  ``{Asymptotic Four Point Functions},''
  \href{http://dx.doi.org/10.1007/JHEP07(2019)082}{{\em JHEP} {\bfseries 07}
  (2019) 082},
\href{http://arxiv.org/abs/1701.04462}{{\ttfamily arXiv:1701.04462 [hep-th]}}.
%%CITATION = ARXIV:1701.04462;%%.

\bibitem{Suzuki:2017ipd}
R.~Suzuki, ``{Refined Counting of Necklaces in One-loop $\mathcal{N}=4$ SYM},''
  \href{http://dx.doi.org/10.1007/JHEP06(2017)055}{{\em JHEP} {\bfseries 06}
  (2017) 055},
\href{http://arxiv.org/abs/1703.05798}{{\ttfamily arXiv:1703.05798 [hep-th]}}.
%%CITATION = ARXIV:1703.05798;%%.

\bibitem{Ryan:2018fyo}
P.~Ryan and D.~Volin, ``{Separated variables and wave functions for rational
  gl(N) spin chains in the companion twist frame},''
  \href{http://dx.doi.org/10.1063/1.5085387}{{\em J. Math. Phys.} {\bfseries
  60} no.~3, (2019) 032701},
\href{http://arxiv.org/abs/1810.10996}{{\ttfamily arXiv:1810.10996 [math-ph]}}.
%%CITATION = ARXIV:1810.10996;%%.

\bibitem{Coronado:2018ypq}
F.~Coronado, ``{Perturbative four-point functions in planar $ \mathcal{N}=4 $
  SYM from hexagonalization},''
  \href{http://dx.doi.org/10.1007/JHEP01(2019)056}{{\em JHEP} {\bfseries 01}
  (2019) 056},
\href{http://arxiv.org/abs/1811.00467}{{\ttfamily arXiv:1811.00467 [hep-th]}}.
%%CITATION = ARXIV:1811.00467;%%.

\bibitem{Jacobsen:2018pjt}
J.~L. Jacobsen, Y.~Jiang, and Y.~Zhang, ``{Torus partition function of the
  six-vertex model from algebraic geometry},''
  \href{http://dx.doi.org/10.1007/JHEP03(2019)152}{{\em JHEP} {\bfseries 03}
  (2019) 152},
\href{http://arxiv.org/abs/1812.00447}{{\ttfamily arXiv:1812.00447 [hep-th]}}.
%%CITATION = ARXIV:1812.00447;%%.

\bibitem{Bajnok:2020xoz}
Z.~Bajnok, J.~L. Jacobsen, Y.~Jiang, R.~I. Nepomechie, and Y.~Zhang,
  ``{Cylinder partition function of the 6-vertex model from algebraic
  geometry},'' \href{http://dx.doi.org/10.1007/JHEP06(2020)169}{{\em JHEP}
  {\bfseries 06} (2020) 169}, \href{http://arxiv.org/abs/2002.09019}{{\ttfamily
  arXiv:2002.09019 [hep-th]}}.

\bibitem{Kuniba:2010ir}
A.~Kuniba, T.~Nakanishi, and J.~Suzuki, ``{T-systems and Y-systems in
  integrable systems},''
  \href{http://dx.doi.org/10.1088/1751-8113/44/10/103001}{{\em J. Phys.}
  {\bfseries A44} (2011) 103001},
\href{http://arxiv.org/abs/1010.1344}{{\ttfamily arXiv:1010.1344 [hep-th]}}.
%%CITATION = ARXIV:1010.1344;%%.

\bibitem{Mukhin2002SolutionsTT}
E.~Mukhin and A.~Varchenko, ``{Solutions to the XXX type Bethe ansatz equations
  and flag varieties},'' {\em Central Eur. J. Math.} {\bfseries 1} (2002)
  238--271, \href{http://arxiv.org/abs/math/0211321}{{\ttfamily
  arXiv:math/0211321 [math.QA]}}.

\bibitem{Mukhin2006QuasipolynomialsAT}
E.~Mukhin and A.~Varchenko, ``{Quasi-polynomials and the Bethe Ansatz},'' {\em
  Geom. Topol. Monogr.} {\bfseries 13} (2008) 385--420,
  \href{http://arxiv.org/abs/math/0604048}{{\ttfamily arXiv:math/0604048
  [math.QA]}}.

\bibitem{2013arXiv1303.1578M}
E.~{Mukhin}, V.~{Tarasov}, and A.~{Varchenko}, ``{Spaces of quasi-exponentials
  and representations of the Yangian Y($gl_N$)},'' {\em Transform. Groups}
  {\bfseries 19} (2014) 861--885,
  \href{http://arxiv.org/abs/1303.1578}{{\ttfamily arXiv:1303.1578 [math.AG]}}.

\bibitem{2013JPhCS.411a2020L}
J.~R. {Li} and V.~{Tarasov},
  \href{http://dx.doi.org/10.1088/1742-6596/411/1/012020}{``{XXZ-type Bethe
  ansatz equations and quasi-polynomials},''} in {\em J. Phys. Conf. Ser.},
  vol.~411, p.~012020.
\newblock 2013.
\newblock \href{http://arxiv.org/abs/1210.2315}{{\ttfamily arXiv:1210.2315
  [math.QA]}}.

\bibitem{Bajnok:2019zub}
Z.~Bajnok, E.~Granet, J.~L. Jacobsen, and R.~I. Nepomechie, ``{On Generalized
  $Q$-systems},'' \href{http://dx.doi.org/10.1007/JHEP03(2020)177}{{\em JHEP}
  {\bfseries 03} (2020) 177},
\href{http://arxiv.org/abs/1910.07805}{{\ttfamily arXiv:1910.07805 [hep-th]}}.
%%CITATION = ARXIV:1910.07805;%%.

\bibitem{Nepomechie:2019gqt}
R.~I. Nepomechie, ``{Q-systems with boundary parameters},'' {\em J. Phys.}
  {\bfseries A53} (2020) 294001,
\href{http://arxiv.org/abs/1912.12702}{{\ttfamily arXiv:1912.12702 [hep-th]}}.
%%CITATION = ARXIV:1912.12702;%%.

\bibitem{Pronko:1998xa}
G.~P. Pronko and {\relax Yu}.~G. Stroganov, ``{Bethe equations 'on the wrong
  side of equator'},''
  \href{http://dx.doi.org/10.1088/0305-4470/32/12/007}{{\em J. Phys.}
  {\bfseries A32} (1999) 2333--2340},
\href{http://arxiv.org/abs/hep-th/9808153}{{\ttfamily arXiv:hep-th/9808153
  [hep-th]}}.
%%CITATION = HEP-TH/9808153;%%.

\bibitem{Pronko:1999gh}
G.~P. Pronko and {\relax Yu}.~G. Stroganov, ``{The Complex of solutions of the
  nested Bethe ansatz. The A(2) spin chain},''
  \href{http://dx.doi.org/10.1088/0305-4470/33/46/309}{{\em J. Phys.}
  {\bfseries A33} (2000) 8267},
\href{http://arxiv.org/abs/hep-th/9902085}{{\ttfamily arXiv:hep-th/9902085
  [hep-th]}}.
%%CITATION = HEP-TH/9902085;%%.

\bibitem{Krichever:1996qd}
I.~Krichever, O.~Lipan, P.~Wiegmann, and A.~Zabrodin, ``{Quantum integrable
  systems and elliptic solutions of classical discrete nonlinear equations},''
  \href{http://dx.doi.org/10.1007/s002200050165}{{\em Commun. Math. Phys.}
  {\bfseries 188} (1997) 267--304},
\href{http://arxiv.org/abs/hep-th/9604080}{{\ttfamily arXiv:hep-th/9604080
  [hep-th]}}.
%%CITATION = HEP-TH/9604080;%%.

\bibitem{Kulish:1979cr}
P.~P. Kulish and N.~{\relax Yu}. Reshetikhin, ``{Generalized Heisenberg
  ferromagnet and the Gross-Neveu model},'' {\em Sov. Phys. JETP} {\bfseries
  53} (1981) 108--114.
[Zh. Eksp. Teor. Fiz.80,214(1981)].
%%CITATION = SPHJA,53,108;%%.

\bibitem{deVega:1989}
H.~J. de~Vega, ``{Yang-Baxter algebras, integrable theories and quantum
  groups},'' \href{http://dx.doi.org/10.1142/S0217751X89000959}{{\em Int. J.
  Mod. Phys.} {\bfseries A04} (1989) 2371--2463}.

\bibitem{Tsuboi:2009ud}
Z.~Tsuboi, ``{Solutions of the T-system and Baxter equations for supersymmetric
  spin chains},'' \href{http://dx.doi.org/10.1016/j.nuclphysb.2009.08.009}{{\em
  Nucl. Phys.} {\bfseries B826} (2010) 399--455},
\href{http://arxiv.org/abs/0906.2039}{{\ttfamily arXiv:0906.2039 [math-ph]}}.
%%CITATION = ARXIV:0906.2039;%%.

\bibitem{Tsuboi:2011iz}
Z.~Tsuboi, ``{Wronskian solutions of the T, Q and Y-systems related to infinite
  dimensional unitarizable modules of the general linear superalgebra
  gl(M|N)},'' \href{http://dx.doi.org/10.1016/j.nuclphysb.2013.01.007}{{\em
  Nucl. Phys.} {\bfseries B870} (2013) 92--137},
\href{http://arxiv.org/abs/1109.5524}{{\ttfamily arXiv:1109.5524 [hep-th]}}.
%%CITATION = ARXIV:1109.5524;%%.

\bibitem{Mukhin:2009}
E.~Mukhin, V.~Tarasov, and A.~Varchenko, ``{Bethe algebra of homogeneous XXX
  Heisenberg model has simple spectrum},'' {\em Commun. Math. Phys.} {\bfseries
  288} (2009) 1--42, \href{http://arxiv.org/abs/0706.0688}{{\ttfamily
  arXiv:0706.0688 [math]}}.

\bibitem{Tarasov:2018}
V.~Tarasov, ``{Completeness of the Bethe ansatz for the periodic isotropic
  Heisenberg model},'' \href{http://dx.doi.org/10.1142/S0129055X18400184}{{\em
  Rev. Math. Phys.} {\bfseries 30} (2018) 1840018}.

\bibitem{Jimbo:1985ua}
M.~Jimbo, ``{Quantum R Matrix for the Generalized Toda System},''
\href{http://dx.doi.org/10.1007/BF01221646}{{\em Commun. Math. Phys.}
  {\bfseries 102} (1986) 537--547}.
%%CITATION = CMPHA,102,537;%%.

\bibitem{Kulish:1985bj}
P.~P. Kulish, ``{Integrable graded magnets},''
  \href{http://dx.doi.org/10.1007/BF01083770}{{\em J. Sov. Math.} {\bfseries
  35} (1986) 2648--2662}.
[Zap. Nauchn. Semin.145,140(1985)].
%%CITATION = JOSMA,35,2648;%%.

\bibitem{Sklyanin:1988yz}
E.~K. Sklyanin, ``{Boundary Conditions for Integrable Quantum Systems},''
\href{http://dx.doi.org/10.1088/0305-4470/21/10/015}{{\em J. Phys.} {\bfseries
  A21} (1988) 2375}.
%%CITATION = JPAGA,A21,2375;%%.

\bibitem{deVega:1992zd}
H.~J. de~Vega and A.~Gonzalez~Ruiz, ``{Boundary K matrices for the six vertex
  and the n(2n-1) A(n-1) vertex models},''
  \href{http://dx.doi.org/10.1088/0305-4470/26/12/007}{{\em J. Phys.}
  {\bfseries A26} (1993) L519--L524},
\href{http://arxiv.org/abs/hep-th/9211114}{{\ttfamily arXiv:hep-th/9211114
  [hep-th]}}.
%%CITATION = HEP-TH/9211114;%%.

\bibitem{deVega:1994sb}
H.~J. de~Vega and A.~Gonzalez-Ruiz, ``{Exact Bethe Ansatz solution for A(n-1)
  chains with nonSU-q(n) invariant open boundary conditions},''
  \href{http://dx.doi.org/10.1142/S0217732394002069}{{\em Mod. Phys. Lett.}
  {\bfseries A9} (1994) 2207},
\href{http://arxiv.org/abs/hep-th/9404141}{{\ttfamily arXiv:hep-th/9404141
  [hep-th]}}.
%%CITATION = HEP-TH/9404141;%%.

\bibitem{Doikou:1998ek}
A.~Doikou and R.~I. Nepomechie, ``{Duality and quantum algebra symmetry of the
  $A^{(1)}_{N-1}$ open spin chain with diagonal boundary fields},''
  \href{http://dx.doi.org/10.1016/S0550-3213(98)00567-7}{{\em Nucl. Phys.}
  {\bfseries B530} (1998) 641--664},
\href{http://arxiv.org/abs/hep-th/9807065}{{\ttfamily arXiv:hep-th/9807065
  [hep-th]}}.
%%CITATION = HEP-TH/9807065;%%.

\end{thebibliography}

\providecommand{\href}[2]{#2}\begingroup\raggedright\endgroup

\end{document}